\documentstyle[subfigure,wrapfig,sprocl]{article}
\edef\psfigRestoreAt{\catcode`@=\number\catcode`@\relax}
\catcode`\@=11\relax
\newwrite\@unused
\def\typeout#1{{\let\protect\string\immediate\write\@unused{#1}}}
\def\figurepath{./}

\def\@nnil{\@nil}
\def\@empty{}
\def\@psdonoop#1\@@#2#3{}
\def\@psdo#1:=#2\do#3{\edef\@psdotmp{#2}\ifx\@psdotmp\@empty \else
    \expandafter\@psdoloop#2,\@nil,\@nil\@@#1{#3}\fi}
\def\@psdoloop#1,#2,#3\@@#4#5{\def#4{#1}\ifx #4\@nnil \else
       #5\def#4{#2}\ifx #4\@nnil \else#5\@ipsdoloop #3\@@#4{#5}\fi\fi}
\def\@ipsdoloop#1,#2\@@#3#4{\def#3{#1}\ifx #3\@nnil 
       \let\@nextwhile=\@psdonoop \else
      #4\relax\let\@nextwhile=\@ipsdoloop\fi\@nextwhile#2\@@#3{#4}}
\def\@tpsdo#1:=#2\do#3{\xdef\@psdotmp{#2}\ifx\@psdotmp\@empty \else
    \@tpsdoloop#2\@nil\@nil\@@#1{#3}\fi}
\def\@tpsdoloop#1#2\@@#3#4{\def#3{#1}\ifx #3\@nnil 
       \let\@nextwhile=\@psdonoop \else
      #4\relax\let\@nextwhile=\@tpsdoloop\fi\@nextwhile#2\@@#3{#4}}
\newread\ps@stream
\newif\ifnot@eof       
\newif\if@noisy        
\newif\if@atend        
\newif\if@psfile       
{\catcode`\%=12\global\gdef\epsf@start{
\def\epsf@PS{PS}
\def\epsf@getbb#1{%
\openin\ps@stream=#1
\ifeof\ps@stream\typeout{Error, File #1 not found}\else
   {\not@eoftrue \chardef\other=12
    \def\do##1{\catcode`##1=\other}\dospecials \catcode`\ =10
    \loop
       \if@psfile
	  \read\ps@stream to \epsf@fileline
       \else{
	  \obeyspaces
          \read\ps@stream to \epsf@tmp\global\let\epsf@fileline\epsf@tmp}
       \fi
       \ifeof\ps@stream\not@eoffalse\else
       \if@psfile\else
       \expandafter\epsf@test\epsf@fileline:. \\%
       \fi
          \expandafter\epsf@aux\epsf@fileline:. \\%
       \fi
   \ifnot@eof\repeat
   }\closein\ps@stream\fi}%
\long\def\epsf@test#1#2#3:#4\\{\def\epsf@testit{#1#2}
			\ifx\epsf@testit\epsf@start\else
\typeout{Warning! File does not start with `\epsf@start'.  It may not be a PostScript file.}
			\fi
			\@psfiletrue} 
{\catcode`\%=12\global\let\epsf@percent=
\long\def\epsf@aux#1#2:#3\\{\ifx#1\epsf@percent
   \def\epsf@testit{#2}\ifx\epsf@testit\epsf@bblit
	\@atendfalse
        \epsf@atend #3 . \\%
	\if@atend	
	   \if@verbose{
		\typeout{psfig: found `(atend)'; continuing search}
	   }\fi
        \else
        \epsf@grab #3 . . . \\%
        \not@eoffalse
        \global\no@bbfalse
        \fi
   \fi\fi}%
\def\epsf@grab #1 #2 #3 #4 #5\\{%
   \global\def\epsf@llx{#1}\ifx\epsf@llx\empty
      \epsf@grab #2 #3 #4 #5 .\\\else
   \global\def\epsf@lly{#2}%
   \global\def\epsf@urx{#3}\global\def\epsf@ury{#4}\fi}%
\def\epsf@atendlit{(atend)} 
\def\epsf@atend #1 #2 #3\\{%
   \def\epsf@tmp{#1}\ifx\epsf@tmp\empty
      \epsf@atend #2 #3 .\\\else
   \ifx\epsf@tmp\epsf@atendlit\@atendtrue\fi\fi}

\def\psdraft{
	\def\@psdraft{0}
}
\def\psfull{
	\def\@psdraft{100}
}

\psfull

\newif\if@draftbox
\def\psnodraftbox{
	\@draftboxfalse
}
\@draftboxtrue

\newif\if@prologfile
\newif\if@postlogfile
\def\pssilent{
	\@noisyfalse
}
\def\psnoisy{
	\@noisytrue
}
\psnoisy
\newif\if@bbllx
\newif\if@bblly
\newif\if@bburx
\newif\if@bbury
\newif\if@height
\newif\if@width
\newif\if@rheight
\newif\if@rwidth
\newif\if@clip
\newif\if@verbose
\def\@p@@sclip#1{\@cliptrue}

\def\@p@@sfile#1{\def\@p@sfile{null}%
	        \openin1=#1
		\ifeof1\closein1%
		       \openin1=\figurepath#1
			\ifeof1\typeout{Error, File #1 not found}
			\else\closein1
			    \edef\@p@sfile{\figurepath#1}%
                        \fi%
		 \else\closein1%
		       \def\@p@sfile{#1}%
		 \fi}
\def\@p@@sfigure#1{\def\@p@sfile{null}%
	        \openin1=#1
		\ifeof1\closein1%
		       \openin1=\figurepath#1
			\ifeof1\typeout{Error, File #1 not found}
			\else\closein1
			    \def\@p@sfile{\figurepath#1}%
                        \fi%
		 \else\closein1%
		       \def\@p@sfile{#1}%
		 \fi}

\def\@p@@sbbllx#1{
		\@bbllxtrue
		\dimen100=#1
		\edef\@p@sbbllx{\number\dimen100}
}
\def\@p@@sbblly#1{
		\@bbllytrue
		\dimen100=#1
		\edef\@p@sbblly{\number\dimen100}
}
\def\@p@@sbburx#1{
		\@bburxtrue
		\dimen100=#1
		\edef\@p@sbburx{\number\dimen100}
}
\def\@p@@sbbury#1{
		\@bburytrue
		\dimen100=#1
		\edef\@p@sbbury{\number\dimen100}
}
\def\@p@@sheight#1{
		\@heighttrue
		\dimen100=#1
   		\edef\@p@sheight{\number\dimen100}
}
\def\@p@@swidth#1{
		\@widthtrue
		\dimen100=#1
		\edef\@p@swidth{\number\dimen100}
}
\def\@p@@srheight#1{
		\@rheighttrue
		\dimen100=#1
		\edef\@p@srheight{\number\dimen100}
}
\def\@p@@srwidth#1{
		\@rwidthtrue
		\dimen100=#1
		\edef\@p@srwidth{\number\dimen100}
}
\def\@p@@ssilent#1{ 
		\@verbosefalse
}
\def\@p@@sprolog#1{\@prologfiletrue\def\@prologfileval{#1}}
\def\@p@@spostlog#1{\@postlogfiletrue\def\@postlogfileval{#1}}
\def\@cs@name#1{\csname #1\endcsname}
\def\@setparms#1=#2,{\@cs@name{@p@@s#1}{#2}}
\def\ps@init@parms{
		\@bbllxfalse \@bbllyfalse
		\@bburxfalse \@bburyfalse
		\@heightfalse \@widthfalse
		\@rheightfalse \@rwidthfalse
		\def\@p@sbbllx{}\def\@p@sbblly{}
		\def\@p@sbburx{}\def\@p@sbbury{}
		\def\@p@sheight{}\def\@p@swidth{}
		\def\@p@srheight{}\def\@p@srwidth{}
		\def\@p@sfile{}
		\def\@p@scost{10}
		\def\@sc{}
		\@prologfilefalse
		\@postlogfilefalse
		\@clipfalse
		\if@noisy
			\@verbosetrue
		\else
			\@verbosefalse
		\fi
}
\def\parse@ps@parms#1{
	 	\@psdo\@psfiga:=#1\do
		   {\expandafter\@setparms\@psfiga,}}
\newif\ifno@bb
\def\bb@missing{
	\if@verbose{
		\typeout{psfig: searching \@p@sfile \space  for bounding box}
	}\fi
	\no@bbtrue
	\epsf@getbb{\@p@sfile}
        \ifno@bb \else \bb@cull\epsf@llx\epsf@lly\epsf@urx\epsf@ury\fi
}	
\def\bb@cull#1#2#3#4{
	\dimen100=#1 bp\edef\@p@sbbllx{\number\dimen100}
	\dimen100=#2 bp\edef\@p@sbblly{\number\dimen100}
	\dimen100=#3 bp\edef\@p@sbburx{\number\dimen100}
	\dimen100=#4 bp\edef\@p@sbbury{\number\dimen100}
	\no@bbfalse
}
\def\compute@bb{
		\no@bbfalse
		\if@bbllx \else \no@bbtrue \fi
		\if@bblly \else \no@bbtrue \fi
		\if@bburx \else \no@bbtrue \fi
		\if@bbury \else \no@bbtrue \fi
		\ifno@bb \bb@missing \fi
		\ifno@bb \typeout{FATAL ERROR: no bb supplied or found}
			\no-bb-error
		\fi
		\count203=\@p@sbburx
		\count204=\@p@sbbury
		\advance\count203 by -\@p@sbbllx
		\advance\count204 by -\@p@sbblly
		\edef\@bbw{\number\count203}
		\edef\@bbh{\number\count204}
}
\def\in@hundreds#1#2#3{\count240=#2 \count241=#3
		     \count100=\count240	
		     \divide\count100 by \count241
		     \count101=\count100
		     \multiply\count101 by \count241
		     \advance\count240 by -\count101
		     \multiply\count240 by 10
		     \count101=\count240	
		     \divide\count101 by \count241
		     \count102=\count101
		     \multiply\count102 by \count241
		     \advance\count240 by -\count102
		     \multiply\count240 by 10
		     \count102=\count240	
		     \divide\count102 by \count241
		     \count200=#1\count205=0
		     \count201=\count200
			\multiply\count201 by \count100
		 	\advance\count205 by \count201
		     \count201=\count200
			\divide\count201 by 10
			\multiply\count201 by \count101
			\advance\count205 by \count201
		     \count201=\count200
			\divide\count201 by 100
			\multiply\count201 by \count102
			\advance\count205 by \count201
		     \edef\@result{\number\count205}
}
\def\compute@wfromh{
		\in@hundreds{\@p@sheight}{\@bbw}{\@bbh}
		\edef\@p@swidth{\@result}
}
\def\compute@hfromw{
		\in@hundreds{\@p@swidth}{\@bbh}{\@bbw}
		\edef\@p@sheight{\@result}
}
\def\compute@handw{
		\if@height 
			\if@width
			\else
				\compute@wfromh
			\fi
		\else 
			\if@width
				\compute@hfromw
			\else
				\edef\@p@sheight{\@bbh}
				\edef\@p@swidth{\@bbw}
			\fi
		\fi
}
\def\compute@resv{
		\if@rheight \else \edef\@p@srheight{\@p@sheight} \fi
		\if@rwidth \else \edef\@p@srwidth{\@p@swidth} \fi
}
\def\compute@sizes{
	\compute@bb
	\compute@handw
	\compute@resv
}
\def\psfig#1{\vbox {
	\ps@init@parms
	\parse@ps@parms{#1}
	\compute@sizes
	\ifnum\@p@scost<\@psdraft{
		\if@verbose{
			\typeout{psfig: including \@p@sfile \space }
		}\fi
		\special{ps::[begin] 	\@p@swidth \space \@p@sheight \space
				\@p@sbbllx \space \@p@sbblly \space
				\@p@sbburx \space \@p@sbbury \space
				startTexFig \space }
		\if@clip{
			\if@verbose{
				\typeout{(clip)}
			}\fi
			\special{ps:: doclip \space }
		}\fi
		\if@prologfile
		    \special{ps: plotfile \@prologfileval \space } \fi
		\special{ps: plotfile \@p@sfile \space }
		\if@postlogfile
		    \special{ps: plotfile \@postlogfileval \space } \fi
		\special{ps::[end] endTexFig \space }
		\vbox to \@p@srheight true sp{
			\hbox to \@p@srwidth true sp{
				\hss
			}
		\vss
		}
	}\else{
		\if@draftbox{		
			\hbox{\fbox{\vbox to \@p@srheight true sp{
			\vss
			\hbox to \@p@srwidth true sp{ \hss \@p@sfile \hss }
			\vss
			}}}
		}\else{
			\vbox to \@p@srheight true sp{
			\vss
			\hbox to \@p@srwidth true sp{\hss}
			\vss
			}
		}\fi

	}\fi
}}
\def\psglobal{\typeout{psfig: PSGLOBAL is OBSOLETE; use psprint -m instead}}
\psfigRestoreAt

\bibliographystyle{unsrt}    
\def\Journal#1#2#3#4{{#1} {\bf #2}, #3 (#4)}
\def\NCA{\em Nuovo Cimento}
\def\NIM{\em Nucl. Instrum. Methods}
\def\NIMA{{\em Nucl. Instrum. Methods} A}
\def\NPB{{\em Nucl. Phys.} B}
\def\PLB{{\em Phys. Lett.}  B}
\def\PRL{\em Phys. Rev. Lett.}
\def\PRD{{\em Phys. Rev.} D}
\def\ZPC{{\em Z. Phys.} C}
\def\st{\scriptstyle}
\def\sst{\scriptscriptstyle}
\def\mco{\multicolumn}
\def\be{\begin{equation}}
\def\ee{\end{equation}}
\def\bea{\begin{eqnarray}}
\def\eea{\end{eqnarray}}
\def\ppbar{\mbox{$p\overline{p}$}}
\def\deg{^{\circ}}
\def\absv#1{\ifmmode \mbox{|#1|}
            \else    \mbox{$|$#1$|$}
            \fi}
\def\qb{\ifmmode  {\overline{q}}
            \else \mbox{$\overline{q}$}
            \fi}    
\newcommand{\qqb}    {\mbox{$q\overline{q}$}} 
\newcommand{\mux}    {\mbox{$\mu\mu$}} 
\newcommand{\llb}    {\mbox{$l\overline{l}$}}  
\newcommand{\pbp}    {\mbox{$\overline{p}p$}} 
\newcommand{\mumu}    {\mbox{$\mu^{+}\mu^{-}$}} 
\newcommand{\zp}    {\mbox{$Z ^{\prime}$}} 
\newcommand{\met}   {\mbox{$\not\!\! E_{T}$}}
\newcommand{\sigbzp}      {\mbox{$\sigma(Z ^{\prime}) \cdot {\rm B}_{ll}$}}
\newcommand{\limsb} {\mbox{limit on $\frac{\lsb}{dM_{z^{\prime}}}$}}
\newcommand{\sigbz}      {\mbox{$\sigma(Z^{\prime}) \cdot {\rm B}$}}
\newcommand{\sqT}   {\mbox{$\sqrt{s}~=~1.8$~TeV}}
\begin{document}
\title{A SEARCH FOR NEUTRAL HEAVY VECTOR GAUGE BOSONS IN $\pbp$ COLLISIONS
AT $\sqT$}
\author{Manoj K. Pillai, E. Hayashi, K. Maeshima, C. Grosso-Pilcher}
\author{P. de Barbaro, A. Bodek, B. Kim, W. Sakumoto}
\address{The CDF Collaboration}
\maketitle\abstracts{
A search for a neutral heavy vector gauge boson, $\zp$, was conducted
in $\pbp$ collisions using 110pb$^{-1}$ of data obtained at the Collider
Detector at Fermilab. We present preliminary 95\% CL limits on the 
production of $\zp$ in different models using its $e^{+}e^{-}$ and
$\mumu$ decay modes.
}

The $\zp$ is a heavy neutral vector gauge boson predicted to exist
in several Grand Unified Theories
and extensions of the Standard
Model.~\cite{GGRoss,delAguila}~The $\zp$  
couplings~\cite{Glashow} to dileptons are described by theory, 
but there are no constraints on its mass. 
\begin{wrapfigure}{l}{2in}
\psfig{figure=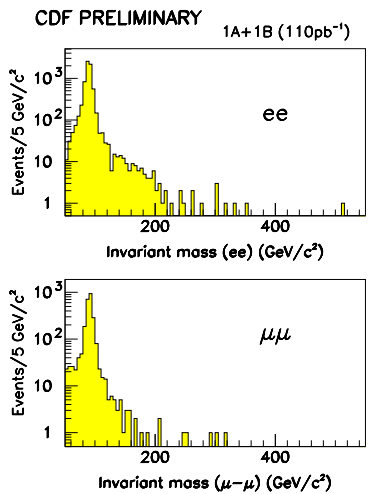,width=2in}
\caption{Distribution of dilepton events as a 
function of invariant mass in 110 pb$^{-1}$ of data taken at CDF.
\label{fig:Mass}}
\end{wrapfigure}
In this paper we describe a direct search for the $\zp$ in its
dielectron and dimuon decay modes at CDF.~\cite{zpPRD}

The calorimeters, the Central Tracking
Chamber (CTC) and the Muon chambers are the principal detector components
used for this search.~\cite{CDFdet}~
Data are collected with a multi-level trigger.
The electron trigger requires a minimum E$_t$ in the calorimeter
with $\approx$ 100\% efficiency. The muon trigger requires a match
between a Central Muon Chamber stub and a high P$_t$ track in the CTC
with $\approx$ 90\% efficiency.

Candidate events are selected by requiring one ``tight'' and one 
``loose'' lepton. Dielectron events are required to have 
one isolated central electron
with E$_t$~$>$~25~GeV/c$^2$ , and P$_t$~$>$~13~GeV/c. 
The second electron could be detected in the Central or in the Plug
region of the detector.
Muons are required to be minimum ionizing
with  P$_t$~$>$~20~GeV/c.
One muon is required to be isolated and detected in the 
Central Muon detector.

We find 7120 dielectron events and 2562 dimuon events. The 
distribution of these events as a function of invariant mass is shown
in Figure~\ref{fig:Mass}. The highest mass $e^{+}e^{-}$ and $\mumu$ events 
have invariant masses of 511~GeV/c$^2$ and 320~GeV/c$^2$ respectively.

\begin{wraptable}{r}{2.3in}
\begin{tabular}{|c|c|c|}\hline
  Mass
 & ee  
 & $\mu\mu$ \\
    (GeV)
    {}
 &  DY + QCD
 &  DY  \\
    {}
 &  pred/data
 &  pred/data  \\
 \hline
  $>$  150    &  68.0/70  &  16.5/17\\
  $>$  200    &  21.7/19  &   6.2/7 \\
  $>$  250    &   8.1/9   &   2.8/4 \\
  $>$  300    &   3.3/6   &   1.4/2 \\    
  $>$  350    &   1.4/2   &   0.7/0 \\
  $>$  400    &   0.7/1   &   0.4/0 \\
  $>$  450    &   0.3/1   &   0.2/0 \\
  $>$  500    &   0.2/1   &   0.1/0 \\
  $>$  550    &   0.1/0   &   0.0/0 \\
  $>$  600    &   0.0/0   &   0.0/0 \\
 \hline
 \end{tabular}
\caption{Predicted number of events from Drell-Yan
and other backgrounds compared with the data.
\label{Dytable}}
\end{wraptable}

Efficiencies of the lepton identification cuts are determined from a sample of  
dileptons from $Z$ decays. The geometric and kinematic acceptance 
were determined from a Monte-Carlo sample of $Z$ and $\zp$ events generated
at different masses. The overall acceptance times efficiency 
rises to a value of 48\% for dielectrons and 20\% for dimuons at 
very high $\zp$ invariant mass.

When the data is compared with the Standard Model expectation and backgrounds, 
we find no significant excess (see~Table~\ref{Dytable}). 
Systematic errors arising from the choice of structure functions and 
P$_t$ distributions are less than 3\%.

The dielectron and dimuon data are combined by assuming that
lepton universality holds for $\zp$ decays. $\zp$ mass limits are 
obtained by comparing the observed data to a superposition of the 
Standard Model prediction and the expected distribution 
from $\zp$ decays using the method of binned likelihood.  

\begin{figure}[here]
\parbox[t]{\textwidth}{
\vspace*{-0.3in}
\mbox{
\subfigure[Limits on $\zp$ production as a function of the $\zp$
mass for a standard model $\zp$]
{\psfig{figure=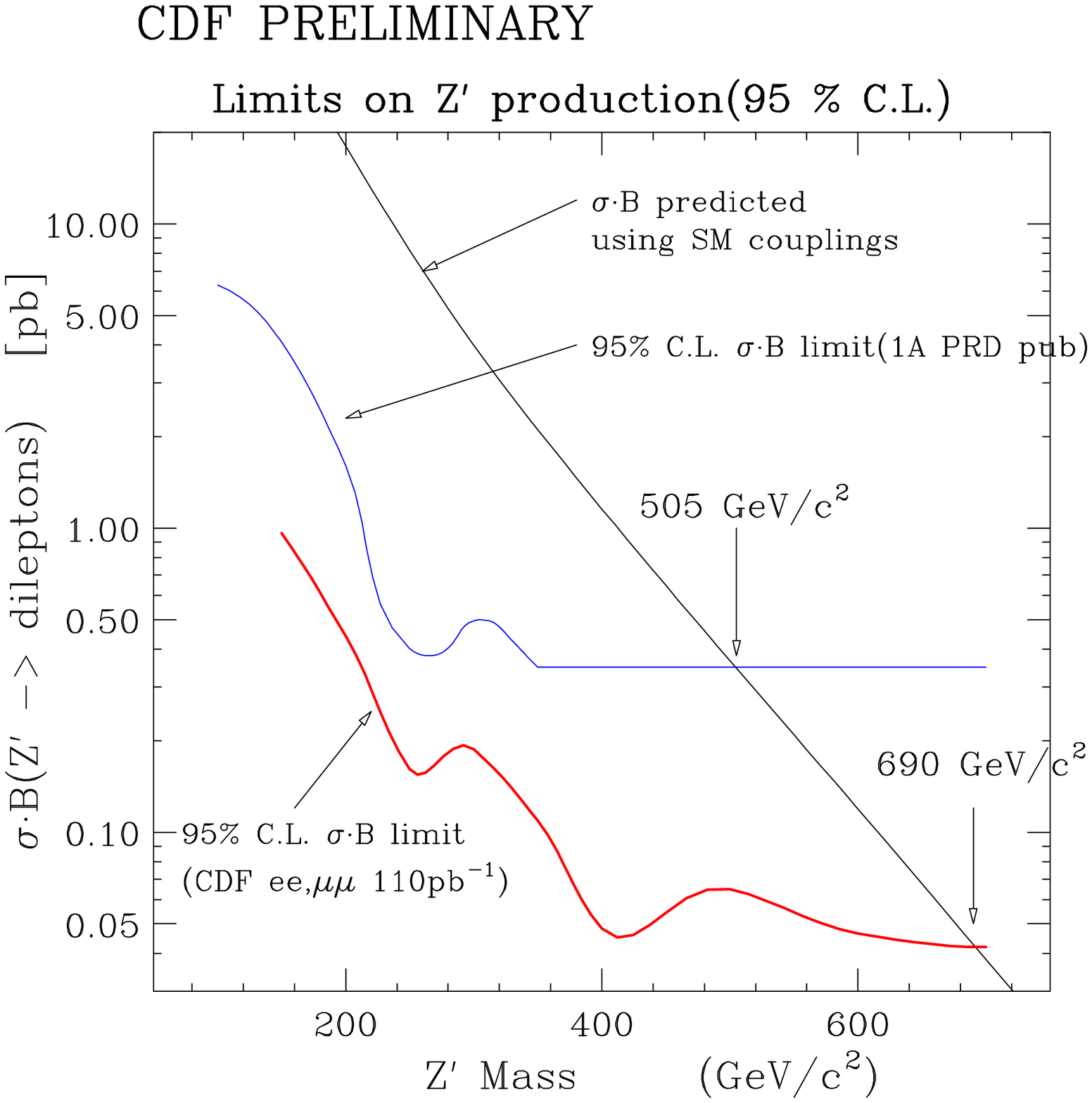,width=2.2in}\label{fig:Std}}\quad
\subfigure[Limits on $\zp$ production as a function of $\zp$ mass  
for several models]{\psfig{figure=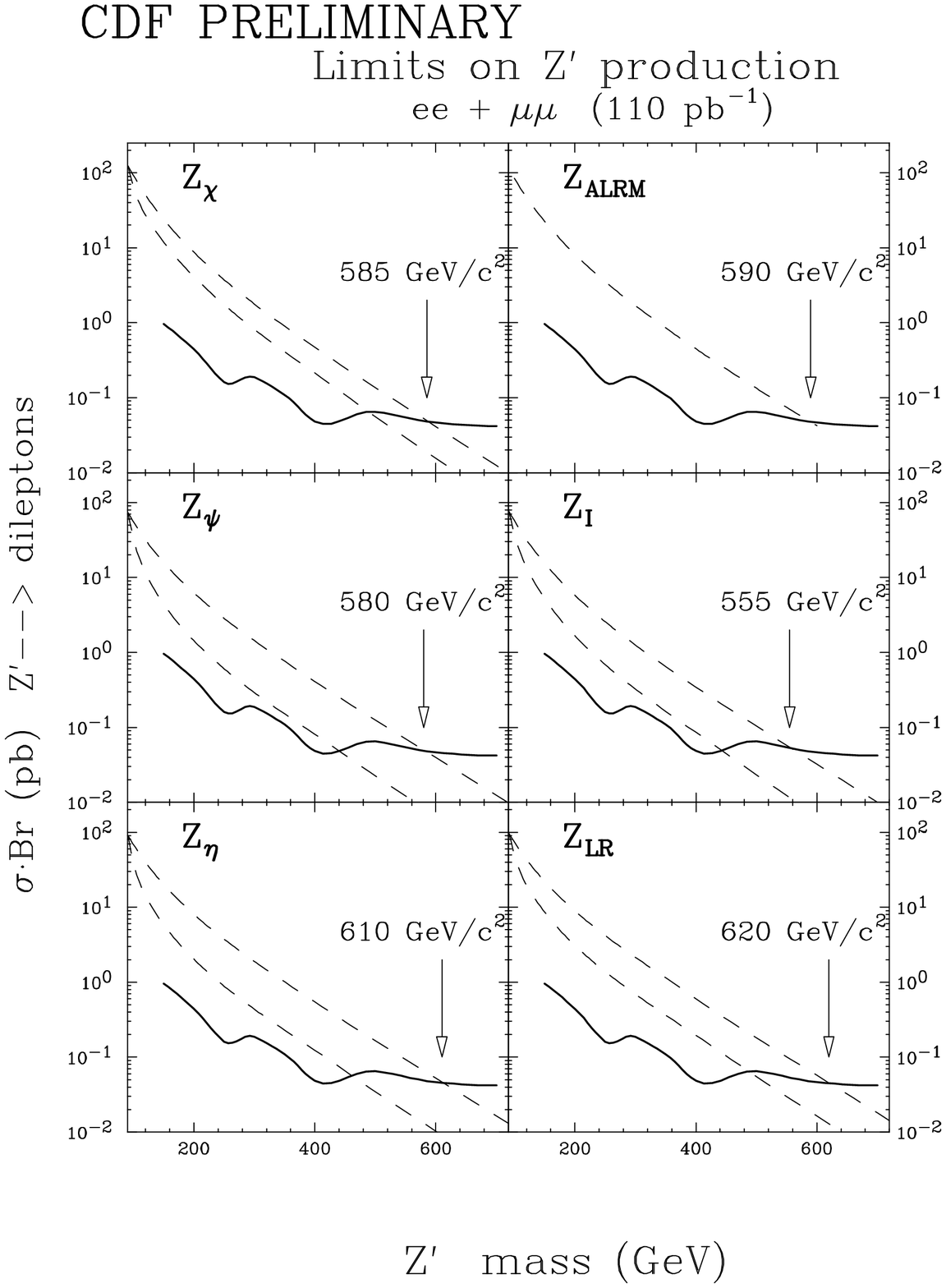,width=2.2in}
\label{fig:Zpmodels}}}
}
\end{figure}
Fig~\ref{fig:Std} shows the 95\% Confidence limit on the cross-section $\times$
branching ratio for a $\zp$ decaying to dileptons, as a function of the $\zp$ mass. For a 
$\zp$ with Standard Model couplings, this translates into a limit of M$_{\zp}$ $>$ 690~GeV/c$^2$ at the
95\% C.L. or a limiting cross-section of 42 fb for M$_{\zp}$ $>$ 650~GeV/c$^2$.
We have also set limits for other interesting theoretical 
models~\cite{delAguila}(see figure~\ref{fig:Zpmodels}).

\section*{Acknowledgments}
We thank the Fermilab staff and the technical staffs of the
participating institutions for their vital contributions.  This work was
supported by the U.S. Department of Energy and National Science Foundation;
the Italian Istituto Nazionale di Fisica Nucleare; the Ministry of Education,
Science and Culture of Japan; the Natural Sciences and Engineering Research
Council of Canada; the National Science Council of the Republic of China;
and the A. P. Sloan Foundation.

\end{document}